\documentclass[aip,twocolumn,letterpaper]{revtex4}

\usepackage{fullpage}

\usepackage{graphics}
\usepackage{epsfig}
\usepackage{xcolor}

\begin{document}
\title{Curl-free magnetic fields for stellarator optimization }
\author{Allen H. Boozer}
\affiliation{Columbia University, New York, NY  10027\\ ahb17@columbia.edu}

\begin{abstract}

This paper describes a new and efficient method of defining an annular region of a curl-free magnetic field with specific physics and coil properties that can be used in stellarator design.  Three statements define the importance: (1) Codes can follow an optimized curl-free initial state to a final full-pressure equilibrium.  The large size of the optimization space of stellarators, approximately fifty externally-produced distributions of magnetic field, makes success in finding a global optimum largely determined by the starting point.  (2)  The design of a stellarator is actually improved when the central region of the plasma has rapid transport with the confinement provided by a surrounding annulus of magnetic surfaces with low transport. (3) The stellarator is unique among all fusion concepts, inertial as well as magnetic, in not using the plasma itself to provide an essential part of its confinement concept.  This permits reliable computational design, which opens a path to faster, cheaper, and more certain achievement of fusion energy. 

\end{abstract}

\date{\today} 
\maketitle

%\tableofcontents

%%%%%%%%%%%%%%%%%%%%%%%%%%%%%%%%%%%%%%%%%%%%%%%%%%%%%%%%%

\section{Introduction}

Lyman Spitzer invented \cite{Spitzer:1951,Spizer:1958} the stellarator in 1951, and it was among the earliest concepts for obtaining the required confinement for fusion energy using magnetic fields.  The stellarator is unique among all fusion concepts, inertial and magnetic, in not requiring any part of the state in which the plasma is confined to be produced by the plasma itself.  Tokamaks require a net plasma current to produce the poloidal magnetic field.  Inertial fusion requires the plasma density $\rho$ times its radius $a$ exceed certain value for ignition.  This is achieved by plasma compression; $\rho a\propto1/a^2$ since $\rho a^3$ is proportional to the mass of the plasma.  The difficulties of producing the required compression in three-dimensional space are discussed in \cite{inertial:2019}. 

Noether's theorem \cite{Noether:1918} provides a strong confining principle for particle motion in an axisymmetric torus.  Unfortunately, the externally controlled state of the stellarator is inconsistent with axisymmetry \cite{Helander:2014}.  A strong breaking of axisymmetry does not require an unacceptably rapid loss of the plasma by the drift of charged particles across the magnetic field lines, but that is the result unless the external magnetic field is carefully designed. 

The benefits of the stellarator are overwhelmed by the problems without careful design optimization.  The required computational power and the mathematical concepts were far beyond what was available in 1951, but they were by the late 1980's.  

Arnulf Schl\"uter had a deep appreciation of the importance stellarator optimization and before his retirement in 1990 lead the Stellaratortheorie group of the Max-Planck-Instituts f\"ur Plasmaphysik, which allowed him to foster the required developments within the group.   By the late 1980's successful design optimizations \cite{Nuhremberg:1988,Grieger:1992} were carried out by J. N\"uhrenberg, P. Merkel, and their collaborators.  They designed the W7-X and the HSX stellarators at that time utilizing somewhat different principles.  These two stellarators are very different in scale, but both experiments \cite{Dinklage:2018,Canik:2007} have demonstrated that careful computational design translates into successful stellarators.  What makes computational design uniquely reliable in stellarators is the independence of the confining state from plasma effects. 

When the reliability of computational design of stellarators is accepted, the whole strategy of fusion energy development can be overturned from what might be called an ITER strategy to a W7-X strategy.  The ITER strategy was to build an experiment that would drive fusion development.  The W7-X strategy was to computationally determine the best conceptual design for an experiment and then use that experiment to validate the design.  Nevertheless, once an experiment is operating, its mission quickly changes into one of driving fusion development.  

The W7-X strategy has three advantages:  (1) The cost of computational design is between 0.1\% and 1\% of the cost of building a major experiment.  (2) Human resources are developed. What can be determined through computational design is dependent upon the creativity of the people in deciding what the goals should be and how they can be achieved.  With modern computers, the effectiveness of computational design has little dependence upon the size of the available computers.  (3) Experiments are costly, but more importantly to the rapid development of fusion, experiments always (a) build in conservatism---even apparently minor changes in design are not possible and therefore remain unstudied---and (b) are built and operated over long periods time.  Success cannot be declared until an experiment proves there has been success, but ideally experiments should only be built to validate a computational design.  

There is now no question that attractive stellarator designs for fusion systems are possible, but what remains unknown is what are the limits on their attractiveness and how rapidly and cheaply fusion could be developed using computational design.  This paper contributes to determining those limits by defining a new, efficient, and more intuitive method of determining stellarator configurations with optimal physics and coil configurations.

The space in which stellarators are optimized has about fifty degrees of freedom---far too many for an optimization code to ensure that a global optimum has been found.  The attractiveness of an optimized state is largely determined by the initial state used in the optimization procedure, which makes constraints on the initial state of great practical importance.  A concept for doing this, which dates from 1964, has had recent development, a Taylor expansion of the magnetic field around the central field line in a toroidal plasma, the magnetic axis \cite{Mericer:1964,Garren:1991,Landreman:2019}.  Unlike axisymmetric systems, the achievement of adequate particle confinement is the primary physics issue in stellarators.  Methods of achieving confinement, quasisymmetry and omnigenity, are discussed in \cite{Boozer:1983,Cary:1997,Landreman:2012}.

Curl-free solutions for the magnetic field are important as initial states for codes that optimize stellarator configurations.    These fields are relatively easy to study, and optimization codes can track an optimum field from a curl-free state to a state with the maximum sustainable plasma pressure.  

%The important changes in the magnetic field produced by the presence of plasma are produced by the current density parallel to the magnetic field $j_{||}$, whether a net or a Pfirsch-Schl\"uter current.  An approximate measure of the importance is $\mu_0j_{||}R/(2\iota B)$, where $\iota$ is the rotational transform and $R$ is the major radius of the torus.  The parallel current is so strong in tokamaks that it determines even the topological properties of the magnetic field.  It can be small in a stellarator, particularly in stellarators of the W7-X type \cite{Grieger:1992}, and produce only small changes in the magnetic field.

While seeking curl-free magnetic fields that have optimal properties, the focus should be the properties in an annulus near the plasma edge--in a region that encloses a toroidal magnetic flux $\psi_t$.  When a curl-free field is optimized in an annulus, then the theory of Laplace's equation states that a curl-free field is robustly defined in the entire region enclosed by the annulus.  

Optimal properties obtained in an annulus will generally not hold in the enclosed region.  But, one may not want them to hold.  It is advantageous for the magnetic surfaces to be broken in the plasma interior and for the transport to be large, although the confining annulus should not be so narrow that the confinement time is too short.  What are the advantages?  (1) A spatially constant pressure $p$ maximizes $\int p^2d^3x$  for a fixed maximum pressure, which maximizes the fusion power.  The confinement time of the plasma is ratio of the total plasma volume to the volume of the annulus longer than the confinement time of the annulus.  (2) Impurities tend to be flushed out more readily the narrower the confinement annulus compared to the total confining volume.  (3) The injection of fuel is easier.  (4)  The optimal 50/50 DT ratio can be maintained in the fusing plasma, which is not trivial when transport coefficients are small in core.

Section \ref{sec:summary} gives a summary of what will be derived in this paper.  Section \ref{sec:tokamaks} discusses the relationship between stellarators and tokamaks.  Section \ref{sec:derivations} derives the expressions for optimizing a curl-free magnetic field in an annulus and how these optimized fields constrain the external magnetic field whether produced by coils, electromagnets, or superconducting tiles. Section \ref{sec:plasma modifications} discusses the modification to the constraints by the plasma.  Section \ref{discussion} is a discussion of the results.  Appendix \ref{sec:examples} derives traditional formulas for the rotational transform of stellarators as examples of the new formalism.

%%%%%%%%%%%%%%%%%%%%%%%%%%%%%%%%%%%%%%%%%%%%%%%%%%

\section{Summary of results  \label{sec:summary}}

This paper develops a method for optimizing curl-free magnetic fields in the vicinity of a toroidal surface $\vec{x}_s(\theta,\varphi)$, where $\theta$ and $\varphi$ are the poloidal and the toroidal angles of Boozer coordinates \cite{Boozer:1981,Boozer:2004,Helander:2014}.  The functional form $\vec{x}_s(\theta,\varphi)$ on a single magnetic surface determines the properties of a curl-free magnetic field.  Two of the features of this field on the $\vec{x}_s$ surface are the local rotational transform $\iota_s$ and the magnetic field strength $B_s(\theta,\varphi)$ in Boozer coordinates, which determines the confinement properties of particle drift trajectories \cite{Boozer:1983,Boozer:2004,Helander:2014}.  For example, the field is quasi-symmetric when $B_s$ has the functional form $B_s(\theta-N\varphi)$, where $N$ is an integer, which implies the Hamiltonian for particle drift trajectories has a confining invariant.   The Hamiltonian for the drift trajectories has a symmetry consistent with Noether's theorem even though the Hamiltonian for the exact particle motion does not.

The rotational transform $\iota_s$ and the magnetic field strength $B_s(\theta,\varphi)$ on the surface are determined by two scalar equations, Equations (\ref{iota-R-Z}) and (\ref{R_s eq}).  Three functions $R(\theta,\varphi)$, $Z(\theta,\varphi)$, and $\omega(\theta,\varphi)$ are available to satisfy these two constraints.  This is consistent with the statement in Garren and Boozer \cite{Garren:1991} that there is sufficient freedom within an expansion around a magnetic axis to achieve exact quasi-symmetry on one magnetic surface.  The existence of three functions for satisfying two constraints implies freedom remains for optimizing the externally produced magnetic field, which is determined by a function of the two angular coordinates of a torus.  The optimization of the external field includes the choice that gives optimal coils and properties in the region enclosed by $\vec{x}_s(\theta,\varphi)$.

The magnetic field on the surface $\vec{x}_s(\theta,\varphi)$ is determined by $\vec{x}_s(\theta,\varphi)$ and a constant, 
\begin{equation}
G_0=\frac{1}{\mu_0}\oint \vec{B}\cdot \frac{\partial x_s}{\partial\varphi}d\varphi,
\end{equation} 
which is the number of Amperes of poloidal current that is flowing in the coils.  Given  $\vec{x}_s(\theta,\varphi)$ and $G_0$, the magnetic field can be determined throughout its curl-free region by using the expression for a general magnetic field as a toroidal field $\mu_0G_0/2\pi R$ plus the sum of magnetic-field distributions that can be produced by magnetic dipoles mounted on a distant enclosing toroidal surface called a coil surface.  These distributions should be numbered by the efficiency \cite{Boozer:NF3D,Landreman:2016} with which currents on the coil surface can produce fields on a toroidal surface similar to $\vec{x}_s(\theta,\varphi)$.  This efficiency drops exponentially with the integer $j$ that numbers the magnetic field distributions; $j_{max}\lesssim50$ is required for an attractive fusion reactor.  The coefficient of each magnetic field distribution is uniquely determined by the requirement that the magnetic field normal to the surface $\vec{x}_s(\theta,\varphi)$ vanish.

%Remarkably, the properties of a magnetic surface defined by $\vec{x}_s(\theta,\varphi)$ and $\vec{x}_s(\theta,N_p\varphi)$ differ by only a simple scaling in $N_p$.  One function $\vec{x}_s(\theta,\varphi)$ gives, through $N_p$ scaling, a whole family of curl-free fields.  Each has a different efficiency of production and different properties away from the toroidal surface $\vec{x}_s(\theta,\varphi)$.  

%The importance of $N_p$ scaling is illustrated by quasihelical helical stellarators.  Garren and Boozer \cite{Garren:1991} showed that quasihelicity is broken by terms of order $(r/R)^3$ where $r$ is the minor and $R$ is the major radius, although exact quasihelical symmetry can be imposed on one surface.  The implication is that a function $\vec{x}_s(\theta,N_p\varphi)$ can be found with exact quasihelical symmetry with the symmetry broken on surfaces deep inside that surface by an amount that scales as $1/N_p^3$.  But as discussed, it is preferable for quasisymmetry to be broken far inside a confining annulus, and this can be accomplished by an appropriate choice of $N_p$.

The convergence properties of the efficient magnetic field distributions \cite{Boozer:NF3D,Landreman:2016} give immediate information on the difficulty of finding a coil set that would accurately produce the field.   Once the fitting to efficient field distributions has been carried out, the magnetic field in the region that is occupied by the plasma is known and its properties, such as the location of magnetic surfaces, the profile of the rotational transform, and the quasi-symmetry, can be easily determined.  Large ports on the outboard side of a toroidal plasma are a central issue in the construction and maintenance of fusion systems, and the consistency of a coil concept with such ports can be simply determined and therefore optimized.  Open access to the plasma chamber is of even greater importance for the fast development of fusion systems since it allows efficient testing of first-wall materials and concepts.  

The magnetic field outside the confining annulus is also determined, but each efficient magnetic field distribution increases exponentially in magnitude between the annulus and the coil surface, the higher the index $j$ the more rapid the exponential increase.  Consequently, small changes in the in $\vec{x}_s$ can produce large changes in the magnetic field between the annulus and the coil surface.  This sensitivity provides freedom for divertor design.

%%%%%%%%%%%%%%%%%%%%%%%%%%%%%%%%%%%%%%%%%%%%%%%%%%%%%%%%%%%%%%%%

\section{Comparison with tokamaks \label{sec:tokamaks} }

Since 1968, magnetic fusion research has been increasingly focused on the tokamak, which has two fundamental differences from a stellarator: (1) In tokamaks, a critical part of the magnetic field that confines the fusing plasma is produced by the currents in the plasma itself but not in stellarators.   (2) The primary control of fusion plasmas is in the externally produced magnetic field.  Axisymmetric tokamaks have approximately five independent distributions of external field, which require careful time-dependent control; stellarators have approximately fifty distributions \cite{Boozer:2004}, which offer far more design freedom and do not require careful time dependent control.  

The differences  between tokamaks and stellarators have manifold implications: (1) During tokamak operations, a sudden loss of plasma control, a disruption, is common \cite{Hender:2007} but disruptions are difficult to induce in stellarators.  Disruptions place a high stress on machine components and can produce destructive beams of relativistic electrons.  A reliable method of avoidance must be invented before tokamak fusion becomes practical.  (2)  In tokamaks, the net plasma current, the part of $j_{||}/B$ that is constant on magnetic surfaces, must not only be maintained but also its spatial distribution must be controlled.  This requires complicated and expensive auxiliary systems of large physical dimensions that use a significant fraction of the fusion power.  (3)  Operating limits are placed on the density and temperature of the plasma, such as the Greenwald limit \cite{Greenwald:1988}.  These limits can force less than optimal operations.  (4) The computational design that is standard for stellarators is far more prone to failure due to the self-confinement of tokamak plasmas.  Nevertheless, the successful design of long wavelength, non-axisymmetric magnetic fields for the control of ELM's in tokamaks has been demonstrated \cite{Park:2018}.  Analogous methods could be used to address other tokamak control issues \cite{Boozer:Nature2018}; similar care and subtlety are required as for successful stellarator design.

%%%%%%%%%%%%%%%%%%%%%%%%%%%%%%%%%%%%%%%%%%%%%%%%%%%%%%%%

\section{Curl-free magnetic fields with optimal properties \label{sec:derivations} }

A magnetic field that is curl free throughout a toroidal region can be represented by two forms \cite{Boozer:2004}
\begin{eqnarray}
2\pi\vec{B} &=& \vec{\nabla}\psi_t\times\vec{\nabla}\theta +\iota(\psi_t) \vec{\nabla}\varphi\times\vec{\nabla}\psi_t \label{contravariant}\\
&=& \mu_0G_0 \vec{\nabla}\varphi \label{covariant}
\end{eqnarray}
wherever magnetic surfaces, $\vec{B}\cdot\vec{\nabla}\psi_t=0$, exist.  The toroidal flux enclosed by a magnetic surface $\psi_t$, the poloidal angle $\theta$, and the toroidal angle $\varphi$, which give this form, are called Boozer coordinates  \cite{Helander:2014}, and $\iota$ is the rotational transform.  

The equality of the two representations of the magnetic field, Equations (\ref{contravariant}) and (\ref{covariant}), ensure the field $\vec{B}$ is a curl-free and divergence-free.  Equation (\ref{contravariant}) forces $\vec{B}$ to be divergence free, and Equation (\ref{covariant}) forces $\vec{B}$ to be curl free.   When  $G_0$ is specified and a function $\vec{x}(\psi_t,\theta,\varphi)$ that satisfies the equality is found, then an explicit expression for a magnetic field $\vec{B}$ has been obtained in Boozer coordinates.  Since $\mu_0G_0$ is a constant, the magnetic field could be found by solving Laplace's equation $\nabla^2\varphi=0$.   When the $(\psi_t,\theta,\varphi)$ coordinates are denoted in the conventional form of general coordinate systems, the appendix of \cite{Boozer:NF3D}, as $(\xi^1,\xi^2,\xi^3)$,
\begin{eqnarray}
&& \nabla^2\varphi = \frac{1}{\mathcal{J}}\sum_{ij} \frac{\partial}{\partial \xi^i} \mathcal{J} g^{ij}  \frac{\partial \varphi}{\partial \xi^i},  \label{Laplace}\\
&& g^{ij} = \vec{\nabla}\xi^i \cdot \vec{\nabla}\xi^j \hspace{0.2in}\mbox{and} \\
&&\frac{1}{\mathcal{J}} = (\vec{\nabla}\xi^1 \times \vec{\nabla}\xi^2) \cdot \vec{\nabla}\xi^3.
\end{eqnarray}
The equality of the two forms for the magnetic field, Equations (\ref{contravariant}) and (\ref{covariant}), implies
\begin{eqnarray}
&&\vec{\nabla}\psi_t\cdot\vec{\nabla}\varphi = 0; \\
&&\vec{\nabla}\theta\cdot\vec{\nabla}\varphi=\frac{\iota}{\mathcal{J}}; \\
&&\vec{\nabla}\varphi\cdot\vec{\nabla}\varphi =\frac{1}{\mathcal{J}}; \\
&&\mathcal{J} = \frac{\mu_0G_0}{(2\pi B)^2}, \hspace{0.2in}\mbox{the coordinate Jacobian.}\hspace{0.2in} \label{Jacobian}
\end{eqnarray}
When these three relations are used in Equation (\ref{Laplace}) together with $\partial \varphi(\psi_t,\theta,\varphi)/\partial\psi_t=0$, $\partial \varphi(\psi_t,\theta,\varphi)/\partial\theta=0$, and $\partial \varphi(\psi_t,\theta,\varphi)/\partial\varphi=1$, one finds that $\nabla^2\varphi$ is always zero.

%%%%%%%%%%%%%%%%%%%%%%%%%%%%%%%

\subsection{Derivation of the constraint equations}

The equality of the two representations of the vector $\vec{B}$, Equations (\ref{contravariant}) and (\ref{covariant}), gives three scalar constraint equations---one for each of the vector components.

Dual relations of the theory of general coordinates, the appendix of \cite{Boozer:NF3D}, can be used to write Equation (\ref{contravariant}) in a form in which it is simpler to obtain the scalar components,
\begin{eqnarray}
\vec{\nabla}\psi_t\times\vec{\nabla}\theta &=& \frac{1}{\mathcal{J}}\frac{\partial\vec{x}}{\partial\varphi}; \\
\vec{\nabla}\varphi\times\vec{\nabla}\psi_t  &=& \frac{1}{\mathcal{J}}\frac{\partial\vec{x}}{\partial\theta}, 
\end{eqnarray}
where the Jacobian $\mathcal{J}$ is given in Equation (\ref{Jacobian}).  Once this is done, the equality of Equations (\ref{contravariant}) and (\ref{covariant}) is equivalent to
\begin{eqnarray}
&& \frac{\partial\vec{x}}{\partial\varphi} + \iota(\psi_t) \frac{\partial\vec{x}}{\partial\theta} = \mathcal{R}^2(\psi_t,\theta,\varphi) \vec{\nabla}\varphi,  \mbox{   where   } \label{constraint}\\
&& B=\frac{\mu_0 G_0}{2\pi \mathcal{R}}.
\end{eqnarray}
The function $\mathcal{R}(\psi_t,\theta,\varphi)$ has units of length and behaves somewhat as a local major radius.

The orthogonality relations of general coordinates, the appendix of \cite{Boozer:NF3D},
\begin{equation}
\frac{\partial\vec{x}}{\partial \xi^i}\cdot \vec{\nabla}\xi^j = \delta_{ij},
\end{equation}
can then be used to obtain the scalar components of  Equation (\ref{constraint});
\begin{eqnarray}
&&  \frac{\partial\vec{x}}{\partial\psi_t}\cdot  \frac{\partial\vec{x}}{\partial\varphi} + \iota(\psi_t)  \frac{\partial\vec{x}}{\partial\psi_t}\cdot \frac{\partial\vec{x}}{\partial\theta}=0; \label{psi comp}\\
&&  \frac{\partial\vec{x}}{\partial\theta}\cdot  \frac{\partial\vec{x}}{\partial\varphi} + \iota(\psi_t)  \frac{\partial\vec{x}}{\partial\theta}\cdot \frac{\partial\vec{x}}{\partial\theta}=0; \\
&&  \frac{\partial\vec{x}}{\partial\varphi}\cdot  \frac{\partial\vec{x}}{\partial\varphi} + \iota(\psi_t)  \frac{\partial\vec{x}}{\partial\varphi}\cdot \frac{\partial\vec{x}}{\partial\theta}=\mathcal{R}^2(\psi_t,\theta,\varphi). 
\end{eqnarray}

%%%%%%%%%%%%%%%%%%%%%%%%%%%%%%%%%%%

\subsection{Implications of the constraint equations}

Although there are three constraint equations, only the last two provide important information on the available curl-free magnetic fields.  The last two define a magnetic surface $\vec{x}_s(\theta,\varphi)$, the rotational transform, and the properties of the particle drift orbits on that surface.  In other words, the only important constraint equations are on the function $\vec{x}_s(\theta,\varphi)$:
\begin{eqnarray}
&&  \frac{\partial\vec{x}_s}{\partial\theta}\cdot  \frac{\partial\vec{x}_s}{\partial\varphi} + \iota_s \frac{\partial\vec{x}_s}{\partial\theta}\cdot \frac{\partial\vec{x}_s}{\partial\theta}=0; \label{theta comp}\\
&&  \frac{\partial\vec{x}_s}{\partial\varphi}\cdot  \frac{\partial\vec{x}}{\partial\varphi} + \iota_s  \frac{\partial\vec{x}_s}{\partial\varphi}\cdot \frac{\partial\vec{x}_s}{\partial\theta}=\mathcal{R}_s^2(\theta,\varphi), \label{phi comp}
\end{eqnarray}
where $\iota_s$ is the rotational transform on the magnetic surface and the magnetic field strength on the surface is $B_s(\theta,\varphi)=\mu_0 G_0/(2\pi \mathcal{R}_s)$.

The efficiency with which this magnetic field can be produced by coils is determinate.  In any toroidal region, the externally produced magnetic field can be separated into a pure toroidal field, specified by $G_0$, and an infinite set of magnetic field distributions ordered by the efficiency with which they can be produced by coils at a distance \cite{Boozer:NF3D,Landreman:2016}.  The condition that there be no field normal to the magnetic surface defined by $\vec{x}_s(\theta,\varphi)$ is
\begin{equation}
\vec{B}\cdot\left( \frac{\partial \vec{x}_s}{\partial\theta}\times\frac{\partial \vec{x}_s}{\partial\varphi}\right)=0. \label{B.n=0}
\end{equation}
Once the required external magnetic field distributions are determined, the magnetic field is known throughout the region in which the externally defined magnetic field distributions are valid.  This solution for the magnetic field is robustly valid within the region enclosed by the magnetic surface $\vec{x}_s(\theta,\varphi)$, so the constraint of Equation (\ref{psi comp}) provides no new information.  Outside the $\vec{x}_s(\theta,\varphi)$ surface, the magnetic field becomes exponentially sensitive to small changes in $\vec{x}_s(\theta,\varphi)$; the further outside the greater the exponential sensitivity.  This exponential sensitivity can be advantageous for the freedom it gives in the design of stellarator divertors.

The requirement that there be no normal field to the magnetic surface, Equation (\ref{B.n=0}), involves a balance between the two types of terms.  The non-axisymmetry of $\vec{x}_s(\theta,\varphi)$ couples the toroidal magnetic field determined by $G_0$ and the collection of magnetic field distributions ordered by their efficiency.  %This balance is changed when the number of toroidal periods $N_p$ in $\vec{x}_s(\theta,\varphi)$ is changed though there is no fundamental change in the function $\vec{x}_s(\theta,\varphi)$.  When a magnetic surface is found that has a desirable rotational transform and drift-orbit properties, an infinite number of solutions that have these properties is obtained by varying the number of periods.  Let $\varphi_s\equiv N_p\varphi$, Equations (\ref{theta comp}) and (\ref{phi comp}) are then
%\begin{eqnarray}
%&&  \frac{\partial\vec{x}_s}{\partial\theta}\cdot  \frac{\partial\vec{x}_s}{\partial\varphi_s} + \frac{\iota_s}{N_p} \frac{\partial\vec{x}_s}{\partial\theta}\cdot \frac{\partial\vec{x}_s}{\partial\theta}=0; \label{theta comp}\\
%&&  \frac{\partial\vec{x}_s}{\partial\varphi_s}\cdot  \frac{\partial\vec{x}}{\partial\varphi_s} + \frac{\iota(\psi_t)}{N_p}  \frac{\partial\vec{x}_s}{\partial\varphi}\cdot \frac{\partial\vec{x}_s}{\partial\theta}=\frac{\mathcal{R}_s^2(\theta,\varphi)}{N_p}. \label{phi comp} \hspace{0.2in}
%\end{eqnarray}

%%%%%%%%%%%%%%%%%%%%%%%%%%%
\subsubsection{Constraint for magnetic coordinates}

Equation (\ref{theta comp}) is the only constraint that must be satisfied for $\theta$ and $\varphi$ to be Boozer magnetic coordinates on the surface $\vec{x}_s(\theta,\varphi)$; magnetic field lines lie in that surface and obey the equation $\theta=\theta_0+\iota_s\varphi$.  The rotational transform, which is a constant, is
\begin{equation}
\iota_s \equiv - \frac{\frac{\partial\vec{x}_s}{\partial\theta}\cdot  \frac{\partial\vec{x}_s}{\partial\varphi}}{ \frac{\partial\vec{x}_s}{\partial\theta}\cdot \frac{\partial\vec{x}_s}{\partial\theta}}. \label{transform1}
\end{equation}
A failure of the right-hand side of this equation to be a constant is a failure to achieve exact magnetic coordinates.

%%%%%%%%%%%%%%%%%%%%%%%%%%%%%

\subsubsection{The magnetic field strength variation}

The third constraint, Equation (\ref{phi comp}), is always satisfied by an arbitrary function $\vec{x}_s(\theta,\varphi)$; it determines the function $\mathcal{R}_s(\theta,\varphi)$.   The important question is whether $\mathcal{R}_s(\theta,\varphi)$ has a form that is consistent with well-confined drift trajectories.  This is assured when $\mathcal{R}_s(\theta,\varphi)$ has a quasi-symmetric form, $\mathcal{R}_s(\theta-N\varphi)$ where $N$ is an integer.  Excellent confinement of drift trajectories can also achieved through omnigenity, which gives a more complicated specification of $\mathcal{R}_s(\theta,\varphi)$.

%%%%%%%%%%%%%%%%%%%%%%%%%%%%%%%%%%%%%%%%%%%%%%%%%%%%%%%%%%%%

\subsection{Cylindrical coordinate expressions}

The position associated with an arbitrary point in $(\psi_t,\theta,\varphi)$ coordinates can be defined using ordinary cylindrical coordinates $(R,\zeta,Z)$;
\begin{eqnarray}
&&\vec{x}(\psi_t,\theta,\varphi) = R(\psi_t,\theta,\varphi) \hat{R}(\zeta)+Z(\psi_t,\theta,\varphi)\hat{Z},\hspace{0.2in}  \\
&&\zeta=\varphi+\omega(\psi_t,\theta,\varphi),\\
&&\frac{d\hat{R}}{d\zeta}=\hat{\zeta}, \hspace{0.2in}\mbox{and}\hspace{0.2in} \frac{d\hat{\zeta}}{d\zeta} =-\hat{R}.
\end{eqnarray}
The two tangent vectors of importance for determining $\vec{x}_s(\theta,\varphi)$ are
\begin{eqnarray}
\frac{\partial\vec{x}_s}{\partial\theta} &=& \frac{\partial R}{\partial\theta}\hat{R} + R \frac{\partial\omega}{\partial\theta}\hat{\zeta} + \frac{\partial Z}{\partial\theta}\hat{Z}; \\
\frac{\partial\vec{x}_s}{\partial\varphi} &=& \frac{\partial R}{\partial\varphi}\hat{R} + R\left(1+ \frac{\partial\omega}{\partial\varphi}\right) \hat{\zeta} + \frac{\partial Z}{\partial\varphi}\hat{Z}.
\end{eqnarray}

The surface $\vec{x}_s(\theta,\varphi)$ is specified when the three functions, $R(\theta,\varphi)$, $Z(\theta,\varphi)$, and $\omega(\theta,\varphi)$ are.  In other words, there are three functions that must satisfy only two constraints.

\subsubsection{Constraint equations in cylindrical coordinates}

Equation (\ref{transform1}), which determines the rotational transform and the magnetic coordinates, has the cylindrical coordinate form
\begin{equation}
\iota_s = - \frac{R^2 \frac{\partial\omega}{\partial\theta}\left(1+ \frac{\partial\omega}{\partial\varphi}\right) + \frac{\partial R}{\partial\theta}\frac{\partial R}{\partial\varphi} + \frac{\partial Z}{\partial\theta} \frac{\partial Z}{\partial\varphi}   }{R^2\left( \frac{\partial\omega}{\partial\theta}  \right)^2 + \left(\frac{\partial R}{\partial\theta}\right)^2+ \left( \frac{\partial Z}{\partial\theta} \right)^2    }. \label{iota-R-Z}
\end{equation}
A solution is found when $\iota_s$ is independent of $\theta$ and $\varphi$.

Equations (\ref{theta comp}) and (\ref{phi comp}) give an equation for $\mathcal{R}_s$ and hence the functional form of the magnetic field strength
\begin{eqnarray}
\mathcal{R}_s^2 &=& R^2\left(1+ \frac{\partial\omega}{\partial\varphi}\right)^2+ \left(\frac{\partial R}{\partial\varphi} \right)^2 + \left( \frac{\partial Z}{\partial\varphi} \right)^2 \nonumber \\
&& - \iota^2 \left\{R^2\left( \frac{\partial\omega}{\partial\theta}  \right)^2 +  \left(\frac{\partial R}{\partial\theta}\right)^2 +\left( \frac{\partial Z}{\partial\theta} \right)^2    \right\}.\hspace{0.2in} \label{R_s eq}
\end{eqnarray}

%%%%%%%%%%%%%%%%%%%%%%%%%%%%%%%%%%%%%%

\subsubsection{Conditions on $R(\theta,\varphi)$ and $Z(\theta,\varphi)$ }

Obtaining a desired form for $B_s(\theta,\varphi)$ or equivalently $\mathcal{R}_s(\theta,\varphi)$ places conditions on $R(\theta,\varphi)$ and $Z(\theta,\varphi)$.  Let $R(\theta,\varphi)=R_0 + \tilde{R}(\theta,\varphi)$, where $R_0$ is a constant and $\tilde{R}(\theta,\varphi)$ is a periodic function of $\theta$ and $\varphi$.   The function $\tilde{R}=R_c+R_b$, where $R_c(\theta,\varphi)$ is consistent with the desired form of $\mathcal{R}_s(\theta,\varphi)$ with $\omega=0$ and $R_b(\theta,\varphi)$ is not.

The largest term in Equation (\ref{R_s eq}) for $\mathcal{R}_s$ is the $R^2$ term.  Obtaining a desired form for $B_s$ requires that $\omega=\omega_b+\tilde{\omega}$, where 
\begin{equation}
\frac{\partial \omega_b}{\partial\varphi}=- \frac{R_b}{R_0}.
\end{equation}
The most subtle term to bring into consistency with $\iota_s$ being constant is $R_0^2\partial\omega/\partial\varphi$.  There are two difficulties.  (1) The expression $R_0^2 \partial\omega/\partial\varphi$ cannot require an $\tilde{\omega}$ that is inconsistent with $\partial \omega_b/\partial\varphi =- R_b/R_0$.  (2) The dominant $\omega$ dependence in the equation for the transform is in the form $\partial \omega/\partial \theta$.  Both difficulties limit the forms that $R(\theta,\varphi)$  and $Z(\theta,\varphi)$ can take.

%%%%%%%%%%%%%%%%%%%%%%%%%%%%%%%%%%%%%%%%%%%%%%%%%%%%%%%%%

\section{Plasma modifications of constraints \label{sec:plasma modifications} }

The effect of a scalar pressure plasma is remarkably small on the constraint equations.  The two expressions for the magnetic field can be written \cite{Boozer:2004}
\begin{eqnarray}
2\pi \vec{B} &=& \frac{1}{\mathcal{J}} \frac{\partial\vec{x}}{\partial \varphi} + \frac{\iota(\psi_t)}{\mathcal{J}} \frac{\partial\vec{x}}{\partial \theta};  \label{contra-p} \\
&=& \mu_0 G(\psi_t) \vec{\nabla}\varphi + \mu_0 I(\psi_t) \vec{\nabla}\theta +\beta_* \vec{\nabla}\psi_t, \mbox{   so   } \hspace{0.2in} \label{co-p} \\
\mathcal{J} &=& \frac{\mu_0(G+\iota I)}{(2\pi B)^2},
\end{eqnarray}
where $\beta_*(\psi_t,\theta,\varphi)$ depends on the pressure gradient.  The two constraint equations on the function $\vec{x}_s(\theta,\varphi)$ are
\begin{eqnarray}
&&  \frac{\partial\vec{x}_s}{\partial\theta}\cdot  \frac{\partial\vec{x}_s}{\partial\varphi} + \iota_s \frac{\partial\vec{x}_s}{\partial\theta}\cdot \frac{\partial\vec{x}_s}{\partial\theta}=\frac{I_s}{G_s+\iota_s I} \mathcal{R}_s^2(\theta,\varphi); \label{genl1} \\
&&  \frac{\partial\vec{x}_s}{\partial\varphi}\cdot  \frac{\partial\vec{x}}{\partial\varphi} + \iota_s  \frac{\partial\vec{x}_s}{\partial\varphi}\cdot \frac{\partial\vec{x}_s}{\partial\theta}=\frac{G_s}{G_s+\iota_s I_s}\mathcal{R}_s^2(\theta,\varphi); \hspace{0.3in}\label{genl2} \\
&& B_s(\theta,\varphi) = \frac{\mu_0(G_s+\iota_s I_s)}{2\pi \mathcal{R}_s},
\end{eqnarray}
where $\iota_s$ is the rotational transform on the surface, $G_s$ is the poloidal current outside of the surface, and $I_s$ is the toroidal current enclosed by the surface; both currents are constants.  Equations (\ref{genl1}) and (\ref{genl2})  are only slightly more complicated to solve than the analogous equations for a curl-free magnetic field, Equations (\ref{theta comp}) and (\ref{phi comp}).

When the surface $\vec{x}_s(\theta,\varphi)$ is sufficiently close to the plasma edge that $G_s$ can be approximated by the poloidal current in the coils $G_0$ and $I_s$ can be approximated by the net plasma current $I_p$, the externally produced magnetic field can be determined in a similar way to the curl-free case.  Nevertheless, there are two differences.  (1) An internal toroidal current and and distribution of dipoles will be required on an inner toroidal surface to fit the magnetic field tangential to the surface $\vec{x}_s(\theta,\varphi)$ as well as ensuring there is no magnetic field normal to this surface.  (2)  The determination of the actual magnetic field in the region enclosed by $\vec{x}_s(\theta,\varphi)$ requires the solution for a non-axisymmetric plasma equilibrium, which may not have magnetic surfaces throughout that region.  The determination of this equilibrium is very slow and complicated compared to just evaluating a sum of externally produced curl-free magnetic field distributions, which was all that was required in the curl-free case.  But, it is important to recall that details of the equilibrium in the region enclosed by $\vec{x}_s(\theta,\varphi)$ are of secondary interest in optimization problems.

It would be ideal if the coil currents required no change to accommodate changed plasma conditions.  In tokamaks, $I_s/(G_s+\iota_s I_s)$ is approximately $(a/R)^2/q\sim 4\%$ and is much smaller in standard stellarator designs, so changes in the plasma require little change in the external magnetic field to maintain the optimization of an outer magnetic surface.   Equations (\ref{genl1}) and (\ref{genl2})  for a plasma with pressure are identical to the equations for a curl-free magnetic field, Equations (\ref{theta comp}) and (\ref{phi comp}) when $I_s=0$.

%%%%%%%%%%%%%%%%%%%%%%%%%%%%%%%%%%%%%%%%%%%%%%%%%%%%%%%%%%%%%%%%%%%

\section{Discussion \label{discussion} }

What has been shown is that stellarator optimization, both plasma and coils, can be efficiently studied using three free functions of two variables each: $R(\theta,\varphi)$, 
$Z(\theta,\varphi)$, and $\omega(\theta,\varphi)$.  Two functional degrees of freedom are required to solve Equations (\ref{iota-R-Z}) and (\ref{R_s eq}) to obtain a desirable form for the magnetic field strength on an outer magnetic surface $\vec{x}_s(\theta,\varphi)$.  Choosing the externally produced magnetic field to have a desirable form requires a third constraint in the form of a  function of a poloidal and a toroidal angle.   The three conditions on the three functions are not trivial and may in part be in conflict.  

Appendix \ref{sec:examples} gives examples of the procedure that require only a few Fourier terms, but to achieve what one wishes to achieve in stellarator optimization a large number of Fourier terms must be included in the representation of $R(\theta,\varphi)$, 
$Z(\theta,\varphi)$, and $\omega(\theta,\varphi)$.   Stellarator symmetry can be imposed by a form related to the Garabedian representation,
\begin{equation}
R+iZ = \sum \Delta_{mn} e^{-i\{(m-1)\theta - n\varphi\}},
\end{equation}
where $R$, $Z$, and the $\Delta_{mn}$ are real, and $\omega=\sum \omega_{mn} \sin\Big((m-1)\theta-n\varphi\Big)$.

The primary purpose of the method developed in this paper is the definition of interesting curl-free magnetic fields that can serve initial states for following into full-pressure plasma equilibria using an optimization code.  Nevertheless, as shown in Section \ref{sec:plasma modifications} the plasma modifications appear very limited when the net toroidal plasma current $I_p$ is extremely small compared to the net poloidal coil current $G_0$, as it often is in stellarators.

%%%%%%%%%%%%%%%%%%%%%%%%%%%%%%%%%%%%%%%%%%%%%%%%%%%%%%%%%%%%%%%%%%%%
\vspace{0.2in}

\section*{Acknowledgements}

This material is based upon work supported by grant 601958 within the Simons Foundation collaboration  "Hidden Symmetries and Fusion Energy" and by the U.S. Department of Energy, Office of Science, Office of Fusion Energy Sciences under Award DE-FG02-95ER54333.   The author would like to thank Matt Landreman for a useful discussion.

\vspace{0.2in}

%%%%%%%%%%%%%%%%%%%%%%%%%%%%%%%%%%%%%%%%%%%%%%%%%%%%%%%%%%%%%%%%%%%%%%%%%%%%%%%
%%%%%%%%%%%%%%%%%%%%%%%%%%%%%%%%%%%%%%%%%%%%%%%%%%%%%

\appendix

%%%%%%%%%%%%%%%%%%%%%%%%%%%%%%%%%%%%%%%%%%%%%%%%%%%%%%%%%

\section{Examples \label{sec:examples} }

Enforcing the constancy of the rotational transform while obtaining a magnetic field strength of a particular form, $B_s(\theta,\varphi)$, requires a large number of Fourier terms, which makes non-trivial but simple examples impossible.  One can illustrate the method, however, by enforcing only the constancy of $\iota_s$ and determining an unconstrained $B_s(\theta,\varphi)$. 

While discussing these examples, let
\begin{eqnarray}
T &\equiv&  \frac{\partial R}{\partial\theta}\frac{\partial R}{\partial\varphi} + \frac{\partial Z}{\partial\theta} \frac{\partial Z}{\partial\varphi} \hspace{0.2in}  \mbox{    and  }\\
\rho &\equiv&  \left(\frac{\partial R}{\partial\theta}\right)^2+ \left( \frac{\partial Z}{\partial\theta} \right)^2
\end{eqnarray}

\subsection{Rotating elliptical surfaces}

\begin{eqnarray}
R&=& R_c + a \Big\{ \cos\theta + \Delta \cos(\theta-N\varphi) \Big\}; \\
Z &=& a \Big\{\sin \theta - \Delta \sin(\theta-N\varphi) \Big\}; \\
\nonumber\\
\frac{\partial R}{\partial\theta} &=& - a \Big\{ \sin\theta + \Delta \sin(\theta-N\varphi) \Big\}; \\
\frac{\partial R}{\partial\varphi} &=& Na\Delta \sin(\theta-N\varphi); \\
\nonumber\\
\frac{\partial Z}{\partial\theta} &=&  a \Big\{ \cos\theta- \Delta \cos(\theta-N\varphi) \Big\}; \\
\frac{\partial Z}{\partial\varphi} &=& Na\Delta \cos(\theta-N\varphi).
\end{eqnarray}

The local minor radius $\rho$ and the drive for the rotational transform $T$ are
\begin{eqnarray}
\rho^2 &=& a^2\Big\{1 - 2\Delta \cos(2\theta - N\varphi) + \Delta^2 \Big\} \\
T &=&Na^2\big\{ \Delta \cos(2\theta-N\varphi)-\Delta^2\Big\}.
\end{eqnarray}

The rotational transform is approximately
\begin{eqnarray}
\iota &=& - \frac{Na^2\Big\{ \Delta \cos(2\theta-N\varphi)-\Delta^2\Big\} + R_c^2\frac{\partial\omega}{\partial\theta} }{\rho^2} \\ 
&\approx& \left(N\Big\{\Delta^2 -\Delta \cos(2\theta-N\varphi)\Big\}  - \left(\frac{R_c}{a}\right)^2 \frac{\partial\omega}{\partial\theta}\right) \nonumber \\  
&&\times \Big(1+ 2\Delta \cos^2(2\theta-N\varphi)\Big)\\
&\approx& N\Delta^2  \mbox{    where  } 
\end{eqnarray}
\begin{eqnarray}
\left(\frac{R_c}{a}\right)^2 \frac{\partial\omega}{\partial\theta} &\approx& -N\Delta \cos(2\theta-N\varphi).
\end{eqnarray}

When $R_c/Na>>1$ dominant term in the variation of the magnetic field strength on the surface is $R^2\approx R_c^2 + 2aR_c\Big\{ \cos\theta + \Delta \cos(\theta-N\varphi) \Big\}$, so
\begin{eqnarray} 
\mathcal{R}^2 &\approx& R_c^2 + 2aR_c\Big\{ \cos\theta + \Delta \cos(\theta-N\varphi) \Big\};  \hspace{0.3in}\\
B&\approx& \frac{\mu_0G_0}{R_c}\Big\{ 1-a\cos\theta -a\Delta \cos(\theta-N\varphi) \Big\}. \hspace{0.3in}
\end{eqnarray}

%%%%%%%%%%%%%%%%%%%%%%%%%%%%%%%%%%%%%%%%%%%

\subsection{Helical axis}

The magnetic axis is assumed to have a helical wobble that is a factor $\Delta_h$ shorter than the length of a period, $R_0/N$.
\begin{eqnarray}
R&=& R_0 + \frac{R_0\Delta_h}{N} \cos(N\varphi) + a \cos\theta \\
Z&=& \frac{R_0\Delta_h}{N} \sin(N\varphi) + a \sin\theta \\
\nonumber\\
\frac{\partial R}{\partial\theta} &=&-a\sin\theta; \\
\frac{\partial R}{\partial\varphi} &=& -R_0\Delta_h \sin(N\varphi)  ;\\
\nonumber\\
\frac{\partial Z}{\partial\theta} &=& a \cos\theta; \\
\frac{\partial Z}{\partial\varphi} &=& R_0\Delta_h \cos(N\varphi);\\
\nonumber\\
\rho^2 &=& a^2 \\
T&=&R_0a\Delta_h \cos(\theta-N\varphi).
\end{eqnarray}

The rotational transform is 
\begin{eqnarray}
\iota &\approx& - \frac{R_0a\Delta_h \cos(\theta-N\varphi) + R_0^2\frac{\partial \omega_0}{\partial\theta} + R_0^2 \frac{\partial \omega_0}{\partial\theta}\frac{\partial \omega_0}{\partial\varphi} }{a^2 +R_0^2\left(\frac{\partial \omega_0}{\partial\theta} \right)^2} \nonumber \\    \nonumber\\
&& + \frac{R_0^2\frac{\partial \omega_1}{\partial\theta} + (R^2-R_0^2)\frac{\partial \omega_0}{\partial\theta}}{a^2 +R_0^2\left(\frac{\partial \omega_0}{\partial\theta} \right)^2};\\
\omega_0&=& -\frac{a\Delta_h}{R_0}\sin(\theta-N\varphi);\\
\iota&\approx& (N-1)\frac{\Delta_h^2}{2}.
\end{eqnarray}

$\mathcal{R}^2$ is not quasi-symmetric since $R^2= R_0^2 + 2(R_0^2\Delta_h/N)\cos(N\varphi) + 2 R_0 a\cos\theta+\cdots$ and other terms do not balance either the $\cos(N\varphi)$ or the $\cos\theta$ expressions.

%%%%%%%%%%%%%%%%%%%%%%%%%%%%%%%%%%%%%%%%%%%%%%%%%%%%%%%%%%%%%%%%%%%%%%%%%%%%%%%%%%%%

%\section{Garabedian representation}

%\begin{equation}
%R+iZ = \sum \Delta_{mn} e^{-i\{(m-1)\theta - n\varphi\}}.
%\end{equation}

%%%%%%%%%%%%%%%%%%%%%%%%%%%%%%%%%%%%%%%%%%%%%%%%%%%%%%%%%%%%%%%%%%%%%%%%%%%%%%%%%%%%%%%%%%%%%%%%%%%%%%%%%%%%%%%%%%%%%%%%

\end{document}